%% file: main.tex
\documentclass[conference]{IEEEtran}
\IEEEoverridecommandlockouts
\usepackage{subfig}
\usepackage{algorithm}
\usepackage{algpseudocode}
\usepackage{cite}
\usepackage{amsmath,amssymb,amsfonts}
\usepackage{graphicx}
\usepackage{textcomp}
\usepackage{siunitx} 
\usepackage{multicol}
\usepackage[nolist]{acronym}
\input{math_commands}

\usepackage{tabularx,booktabs}
\usepackage{longtable}
\usepackage{color, colortbl,booktabs}
\usepackage{multirow}
\usepackage{url}

\def\BibTeX{{\rm B\kern-.05em{\sc i\kern-.025em b}\kern-.08em
    T\kern-.1667em\lower.7ex\hbox{E}\kern-.125emX}}

\begin{document}
\input{./acronyms.tex}
\title{Vision-Assisted Digital Twin Creation for mmWave Beam Management}
\author{
    \IEEEauthorblockN{Maximilian Arnold\IEEEauthorrefmark{1}, Bence Major\IEEEauthorrefmark{1}, Fabio Valerio Massoli\IEEEauthorrefmark{1}, Joseph B. Soriaga\IEEEauthorrefmark{2}, Arash Behboodi\IEEEauthorrefmark{1} 
}
    \IEEEauthorblockA{
        \IEEEauthorrefmark{1}Qualcomm Technologies Netherlands B.V.,
        \IEEEauthorrefmark{2}Qualcomm Technologies Inc.\\
    }
}
\maketitle
\vspace{-5mm}
\begin{abstract}
In the context of communication networks, digital twin technology provides a means to replicate the \ac{RF} propagation environment as well as the system behaviour, allowing for a way to optimize  the performance of a deployed system based on simulations. 
One of the key challenges in the application of Digital Twin technology to mmWave systems is the prevalent channel simulators' stringent requirements on the accuracy of the 3D Digital Twin, reducing the feasibility of the technology in real applications.
We propose a practical Digital Twin creation pipeline and a channel simulator, that relies only on a single mounted camera and position information. We demonstrate the performance benefits compared to methods that do not explicitly model the 3D environment, on downstream sub-tasks in beam acquisition, using the real-world dataset of the DeepSense6G challenge.
\end{abstract}

\input{sections/1_introduction}
\input{sections/2_dataset}
\input{sections/3_digital_twin}
\input{sections/4_downstream}
\input{sections/5_experiments}

\input{sections/6_conclusions}
\bibliography{bibliography}
\bibliographystyle{IEEEtran}
\end{document}

%% file: math_commands.tex
\usepackage{amsmath,amsfonts,bm}

\def\eqref#1{equation~\ref{#1}}

\def\1{\bm{1}}

\DeclareMathAlphabet{\mathsfit}{\encodingdefault}{\sfdefault}{m}{sl}
\SetMathAlphabet{\mathsfit}{bold}{\encodingdefault}{\sfdefault}{bx}{n}

\newcommand*{\norm}[1]{\left\|#1\right\|}

\usepackage{todonotes}



%% file: acronyms.tex
\begin{multicols}{2}
\begin{acronym}[WSSUS]
    \acro{3GPP}{3rd Generation Partnership Project}
    \acro{5G}{fifth-generation}
    \acro{6G}{sixth-generation}
    \acro{AoA}{angle-of-arrival}
    \acro{ADC}{analog to digital converter}
    \acro{AFE}{analog front end}
    \acro{AGC}{automatic gain control}
    \acro{AGV}{automated guided vehicle}
    \acro{AMP}{approximate message passing}
    \acro{API}{Application Programming Interface}
    \acro{AWGN}{additive white Gaussian noise}
	\acro{AE}{auto-encoder}
     \acro{BCE}{binary cross-entropy}
    \acro{BER}{bit error rate}
    \acro{BB}{baseband}
    \acro{bpcu}{bits per channel use}
    \acro{BP}{belief propagation}
    \acro{BPSK}{binary phase shift keying}
    \acro{BS}{base station}
    \acro{CB}{codebook}
    \acro{CE}{cross-entropy}
    \acro{CL}{contrastive-loss}
    \acro{CFAR}{Constant-False-Alarm-Rate}
    \acro{CDF}{cumulative distribution function}
    \acro{CFO}{carrier frequency offset}
    \acro{CoSaMP}{compressive sampling matching pursuit}
    \acro{CP}{cyclic prefix}
    \acro{CS}{compressive sensing} 
    \acro{CSI}{channel state information}
    \acro{CNN}{convolutional neural network}
	\acro{DBSCAN}{Density-Based-Spatial-Clustering-of-Applications}
    \acro{DA}{domain adaptation}
    \acro{DBA}{Distance-based Accuracy}
    \acro{DAC}{digital-analog-converter}
    \acro{DC}{direct current}
    \acro{DE}{distance error}
    \acro{DeepL}{deep-learning}
    \acro{DoF}{degree-of-freedom}
    \acro{DFT}{discrete Fourier transformation}
    \acro{DL}{deep learning}
    \acro{DT}{digital twin}
    \acro{DS}{delay spread}
    \acro{DGPS}{Differential Global Positioning Systems}
    \acro{DSP}{digital signal processing}
	\acro{EPS}{Constant-False-Alarm-Rate}
    \acro{ECC}{error-correcting code}
    \acro{ENoB}{effective number of bits}
    \acro{ERP}{effective radiated power}
    \acro{E2E}{end-2-end}
    \acro{EVM}{error vector magnitude}
    \acro{EVD}{eigenvector decomposition}
    \acro{FB}{feedback}
    \acro{FP}{false positive}
    \acro{FN}{false negative}
    \acro{FC}{fully connected}
    \acro{FDD}{frequency division duplexing}
    \acro{FDM}{frequency division multiplexing}
    \acro{FIR}{finite impulse response}
    \acro{FFT}{fast fourier transform}
    \acro{FT}{fine tuning}
    \acro{FPGA}{field programmable gate array}
    \acro{GAN}{Generative adversarial network}
    \acro{GPIO}{general-purpose input/output}
    \acro{GPS}{global positioning system}
    \acro{GPSDO}{GPS disciplined oscillator}
    \acro{GPU}{graphical processing unit}
    \acro{HDF}{Hierarchical Data Format}
    \acro{HDD}{hard decision decoding}
    \acro{IC}{integrated circuit}
    \acro{ICI}{inter-carrier-interference}
    \acro{ISAC}{Integrated Sensing And Communication}
    \acro{I2C}{Inter-Integrated Circuit}
    \acro{ICSP}{in-circuit serial programming}
    \acro{IF}{intermediate frequency}
    \acro{i.i.d.}{independent and identically distributed}
    \acro{IIR}{infinite impulse response}
    \acro{IMU}{inertial measurement unit}
    \acro{IoT}{Internet of Things}
    \acro{IoU}{Intersection over Union}
    \acro{IPS}{indoor positioning system}
    \acro{IR}{infrared}
    \acro{JSDM}{Joint Spatial Division and Multiplexing}
    \acro{LiDAR}{Light Detection And Ranging}
    \acro{LLR}{log-likelihood ratio}
    \acro{LP}{leakage precoder}
    \acro{LMMSE}{Linear Minimum Mean Square Error}
    \acro{LO}{local oscillator}
    \acro{L1}{Layer-1}
    \acro{L2}{Layer-2}
    \acro{LoS}{line of sight}
    \acro{LiDaR}{Light Detection and Ranging}
    \acro{LS}{least squares}
    \acro{LSTM}{long-term short-term memory}
    \acro{LTE}{Long Term Evolution}
    \acro{LTI}{linear time invariant}
    \acro{LTV}{linear time variant}
    \acro{MAP}{maximum a posteriori}
    \acro{ML}{maximum likelihood}
    \acro{MSE}{mean squared error}
    \acro{mmWave}{millimetre Wave}
    \acro{MUSIC}{Multiple Signal Classification}
    \acro{NN}{Neural Network}
    \acro{NERF}{NEural Radiance Fields}
    \acro{MLP}{multilayer perceptron}
    \acro{NNI}{Neural Network Intelligence}
    \acro{NLoS}{non-line of sight}
    \acro{KNN}{k-nearest neighbors}
    \acro{KPI}{key performance indicator}
    \acro{OFDM}{orthogonal frequency division multiplex}
    \acro{RADAR}{Radio Detection And Ranging}
    \acro{RGB}{Red-Green-Blue}
    \acro{ReLU}{rectified linear unit}
    \acro{RF}{radio frequency}
    \acro{RMS-DS}{Root Mean Square - Delay Spread}
    \acro{RNN}{recurrent neuronal network}
    \acro{RSSI}{received signal strength indicator}
    \acro{R-ZF}{regularized zero-forcing}
    \acro{SDD}{soft decision decoding}
    \acro{SDR}{software defined radio}
    \acro{SE}{spectral efficiency}
    \acro{SFO}{sampling frequency offset}
    \acro{STO}{sampling time offset}
    \acro{SLAM}{Simultaneous Localization and Mapping}
    \acro{SGD}{stochastic gradient descent}
    \acro{SISO}{single input single output}
    \acro{SINR}{signal-to-interference-and-noise-ratio}
    \acro{SIR}{signal-to-interference-ratio}
    \acro{SLNR}{signal-to-leakage-and-noise ratio}
    \acro{SNR}{signal-to-noise-ratio}
    \acro{SP}{subspace}
    \acro{SQR}{signal-to-quantization-noise-ratio}
    \acro{SQNR}{signal-to-quantization-noise-ratio}
    \acro{SVD}{singular value decomposition}
    \acro{SU}{single-user}
    \acro{TDD}{time division duplexing}
    \acro{TRIPS}{time-reversal IPS}
    \acro{TRP}{transmission and reception point}
    \acro{TP}{true positive}
    \acro{TN}{true negativ}
    \acro{UE}{user equipment}
    \acro{UL}{uplink}
    \acro{ULA}{uniform linear array}
    \acro{URLLC}{ultra-reliable low-latency communication}
    \acro{US}{uncorrelated scattering}
    \acro{USRP}{universal software radio peripheral}
    \acro{UWB}{ultra-wideband}
    \acro{WiFi}{Wireless Fidelity}
    \acro{WSS}{wide sense stationary}
    \acro{WSSUS}{wide sense stationary uncorrelated scattering}
	\acro{YOLO}{You-Only-Look-Once}
    \acro{ZF}{zero forcing}
\end{acronym}
\end{multicols}

%% file: sections/1_introduction.tex
\section{Introduction}\label{sec:introduction}
In recent years, \ac{DT} technology has emerged as a revolutionary approach to enhance the design, operation, and optimization in various disciplines\cite{8901113}. By creating virtual replicas of real-world assets, \acp{DT} provide an interactive and dynamic platform for analyzing performance, predicting behavior, and facilitating decision-making processes.

In the context of communication networks, \ac{DT}s are envisioned to enhance network efficiency by enabling more accurate and efficient optimization of various network parameters both during deployment (e.g.\ base stations positions/rotations\cite{orekondy2023winert,choi2023withray,wireless_insite}) and also during system operation. One of the main beneficiary use-cases of supplementary \ac{DT} information is \ac{mmWave} beam management, due to the challenges of establishing and maintaining directional links.

Due to the wave propagation characteristics at \ac{mmWave} frequencies, (specular behavior, strong attenuation and short coherence length) it is essential to simulate the propagation of electromagnetic waves by tracing the trajectories of individual rays and accounting for various phenomena, including reflection, diffraction, and scattering\cite{sionna-rt}. This necessitates the  modeling of real-world complexities of the environment with high degree of fidelity. The most prominent representation used for this purpose are 3D meshes and assigned material properties for the given deployment frequency. 

While there is prior art for \ac{mmWave} beam management using 3D representations\cite{jiang_digital_2023,zhang2023digital}, these approaches rely on the existence of 3D meshes with high fidelity, which greatly increases the cost of deployment, as they require precise \emph{digitalisation} of the environment. Moreover, for a practical solution, the same information needs to be acquired also for dynamic objects in real-time.

A possible solution to acquire real-time 3D information is Deep Learning based Computer Vision, using, for example, an \ac{RGB}-camera co-located with the base-station. These techniques can reach high reconstruction quality in a multi-view setting\cite{nerf2021}, which is not practical due to large deployment/maintaining effort. Single-view scene reconstruction solutions, on the other hand, lack accuracy of the estimated surface normals\cite{yu2022monosdf}.

An alternative approach is proposed by \cite{alrabeiah_millimeter_2020}, who maps a camera image (using a neural network) \emph{directly} to the selected beam index, without explicit modeling of the 3D environment in an intermediate step. We refer to such approaches that omit this intermediate step as \emph{end-to-end approaches}. These solutions, being black-box and hard to interpret, do not leverage \ac{RF} domain knowledge and can base their decisions on irrelevant data patterns, and thereby harm generalizability.

This poses the question: Is there a practical \emph{3D reconstruction} technique, in conjunction with a robust \emph{channel simulator}, that together enable enhanced communication? 

\begin{figure}[t]
     \includegraphics{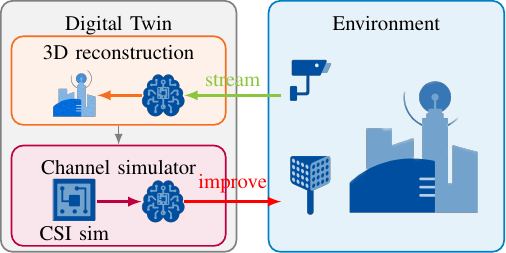}
     \caption{Pipeline to automatically create a 3D environment, simulate the channel state and improve the communication network.}
    \label{fig:sys-view}
\vspace{-5mm}  
\end{figure}

Summarizing our contributions:
\begin{itemize}
\item We propose a \ac{DT} framework that automatically (without \textbf{manual} intervention) reconstructs the 3D environment using a stationary camera, and then estimates the channel state (see Fig.~\ref{fig:sys-view}).
\item To demonstrate the use of \ac{DT} for beam management, we propose and implement solutions for downstream sub-problems in beam acquisition.
\item On a real-world dataset, we demonstrate that our \ac{DT}-based solution outperforms the state-of-the-art end-to-end methods\cite{avatar2022,jiang_digital_2023}, and we analyze the differences in explainability.
\end{itemize}

%% file: sections/2_dataset.tex
\section{Dataset} \label{sec:dataset}
DeepSense \ac{6G}~\cite{DeepSense} is a comprehensive and expansive dataset curated specifically for the research and development of \ac{6G} technologies.
Inspired by the multi-modal beam prediction challenge, and to compare the performance of our proposed system, we are using the four scenarios (Scenario 31-34) from the challenge.

The experimental setup of these specific scenarios consists of a \ac{mmWave} \ac{BS} with a mounted camera with a horizontal field of view of \SI{110}{\degree} and resolution of 960x540 pixels.  The \ac{BS} is equipped with a 16-element \ac{ULA} antenna using a \ac{DFT} \ac{L2} \ac{CB} with size 64. The beam profiles are measured in anechoic chamber and  cover the angular domain from \SI{-50}{\degree} until \SI{50}{\degree} with \SI{1}{\degree} steps. We artificially create the \ac{L1} \ac{CB} by using a \ac{CB} with size 6 covering the same angular domain as the \ac{L2} \ac{CB}. The \ac{UE} uses an omni-directional beam with the same type of a 16-element \ac{ULA}. For each beam, the measured received power is given excluding phase information.

The UE location is given by \ac{DGPS}.  We note that the position accuracy is an order of magnitude better than regular off-the-shelf devices, yet in a deployment case, the known \ac{BS} position at the receiver side can be leveraged to correct the \ac{GPS} position.
For more information about the other scenarios, we refer the interested reader to \cite{DeepSense}.

\subsection{Reconstruction of angular power profile}\label{sec:reconstruction_angle_profile}
As the dataset provides power measurements using analogue beamforming, it prevents us to test different downstream tasks on the data. To overcome this limitation, we reconstruct an angular power profile and use it as a proxy for the full channel-state-information. 

The reconstruction is done as follows:    
\begin{equation}
\mathbf{r}_{\alpha} = \left|\sum_{k \in \mathcal{K}} y_k\mathbf{b}_{k}\right| \in \mathbb{R}_{+}^{N_{\text{angles}}}
\end{equation}
where $\mathcal{K}$ is the set of the top-$K$ measurements, $y_k \in \mathbb{R}_{+}$ the magnitude of the measured received power for beam $k$, $\mathbf{b}_{k} \in \mathbb{R}_{+}^{N_{\text{angles}}}$ is the beam profile of beam $k$, measured in an anechoic chamber. To compensate for noise, we use a larger number of measurements ($K=16$). Here, we have a spatial sampling of \SI{1}{\degree} with coverage of $N_\text{angles}=180$. 

Using the reconstructed angular power profile, we can simulate the received power of any beam given by its beam profile $\mathbf{\tilde{b}}_{k} \in \mathbb{R}_{+}^{N_{\text{angles}}}$ 
\begin{equation}\label{eq:powerrecon}
\hat{y}_k = \mathbf{\tilde{b}}_{k}^{
\text{T}} \mathbf{r}_{\alpha},
\end{equation}
to estimate its received power $\hat{y}_k \in \mathbb{R}_{+}$.
\begin{figure}[t]
     \includegraphics{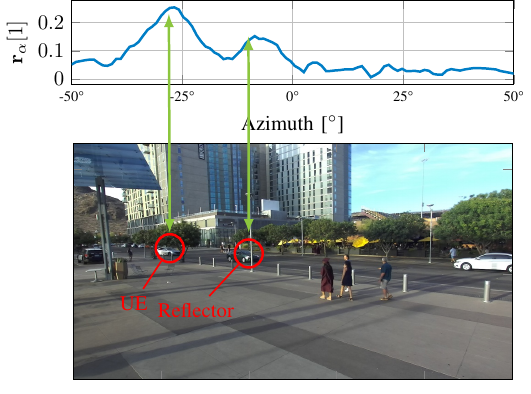}
     \vspace{-6mm}
     \caption{Example for a reconstructed angular power profile comparing to the camera image and identifying useful reflectors. Note the camera distortion due to not-available checker board calibration matrix.}
\label{fig:angle_example}
\vspace{-4mm}  
\end{figure}

Fig.~\ref{fig:angle_example} depicts the physical meaning of the reconstructed angular power profile, where the main energy is pointing to the \ac{LoS} path of the \ac{UE}, while part of the energy is reflected over other objects. Overall, the filtering keeps the original measured signals and suppresses a large part of the noisy measurements. Since the angular profile is a coarse proxy for the channel state information, we cannot use it to simulate the 5G beam management procedure.

\begin{figure*}[t]
    \centering
    \subfloat[\centering Scenario 31]{{\includegraphics[width=0.24\textwidth,height=4cm]{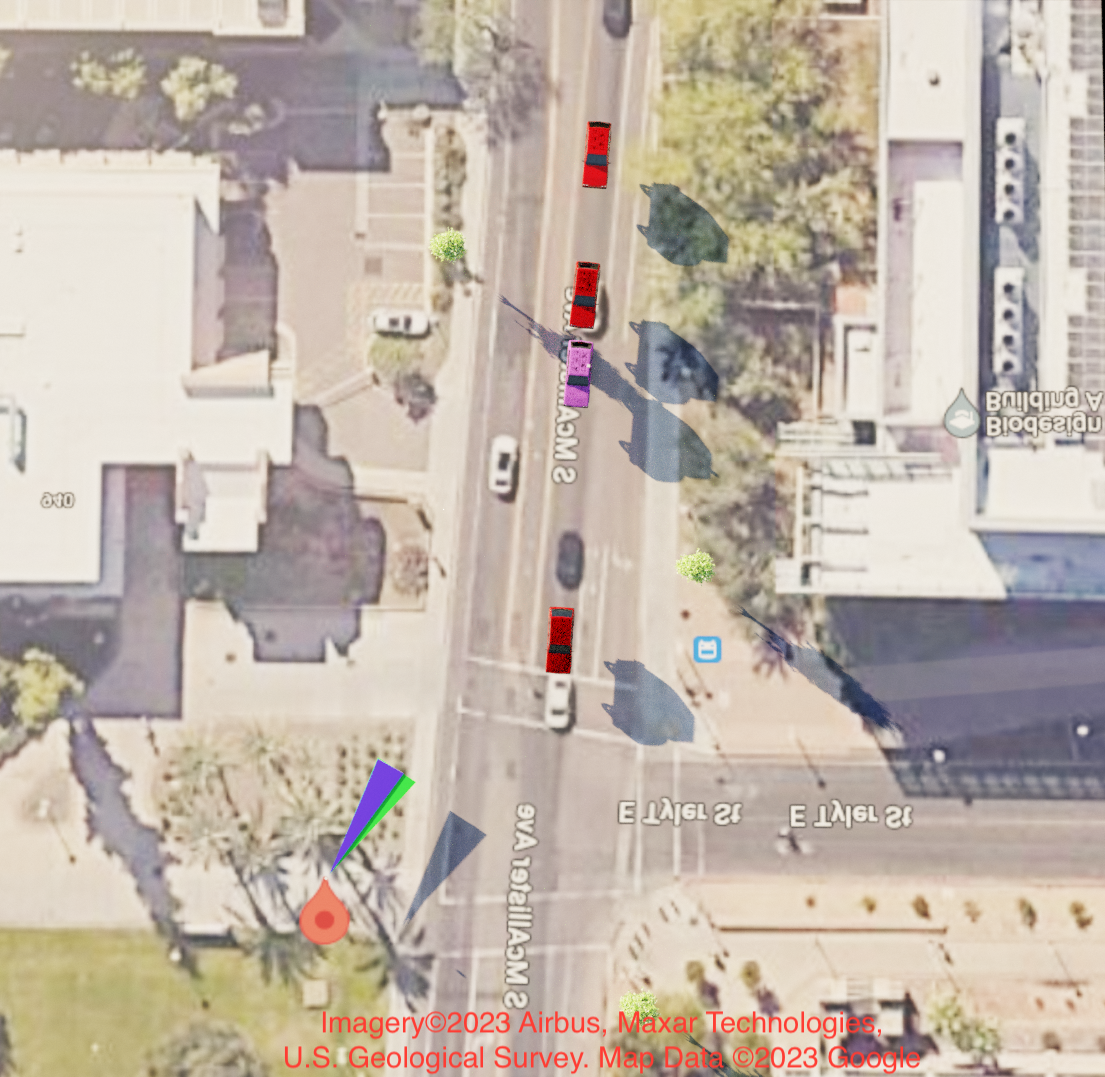}}}%
    \quad \hspace*{-0.9em}
    \subfloat[\centering Scenario 32]{{\includegraphics[width=0.24\textwidth,height=4cm]{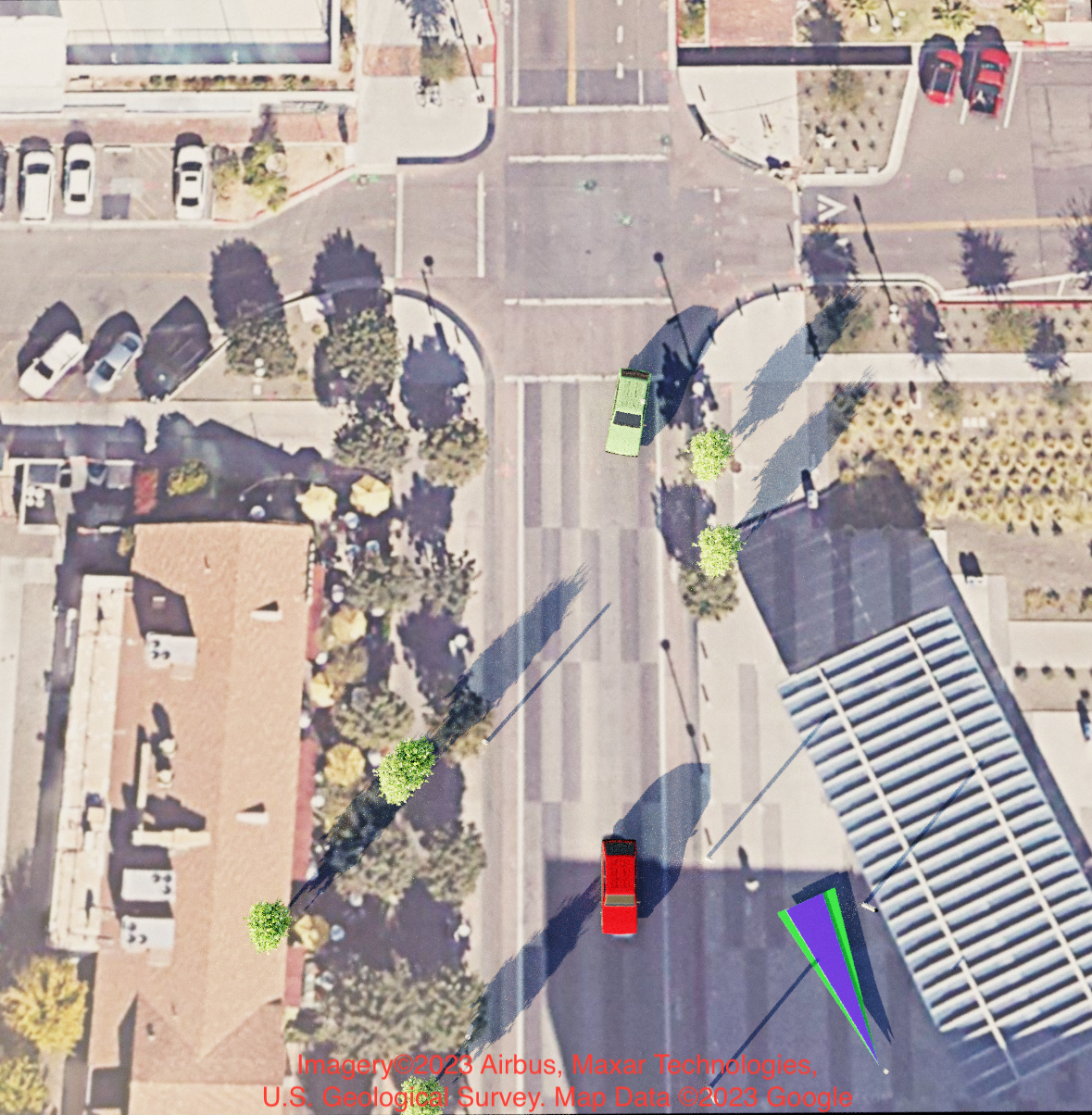}}}%
    \quad \hspace*{-0.9em}
    \subfloat[\centering Scenario 33]{{\includegraphics[width=0.24\textwidth,height=4cm]{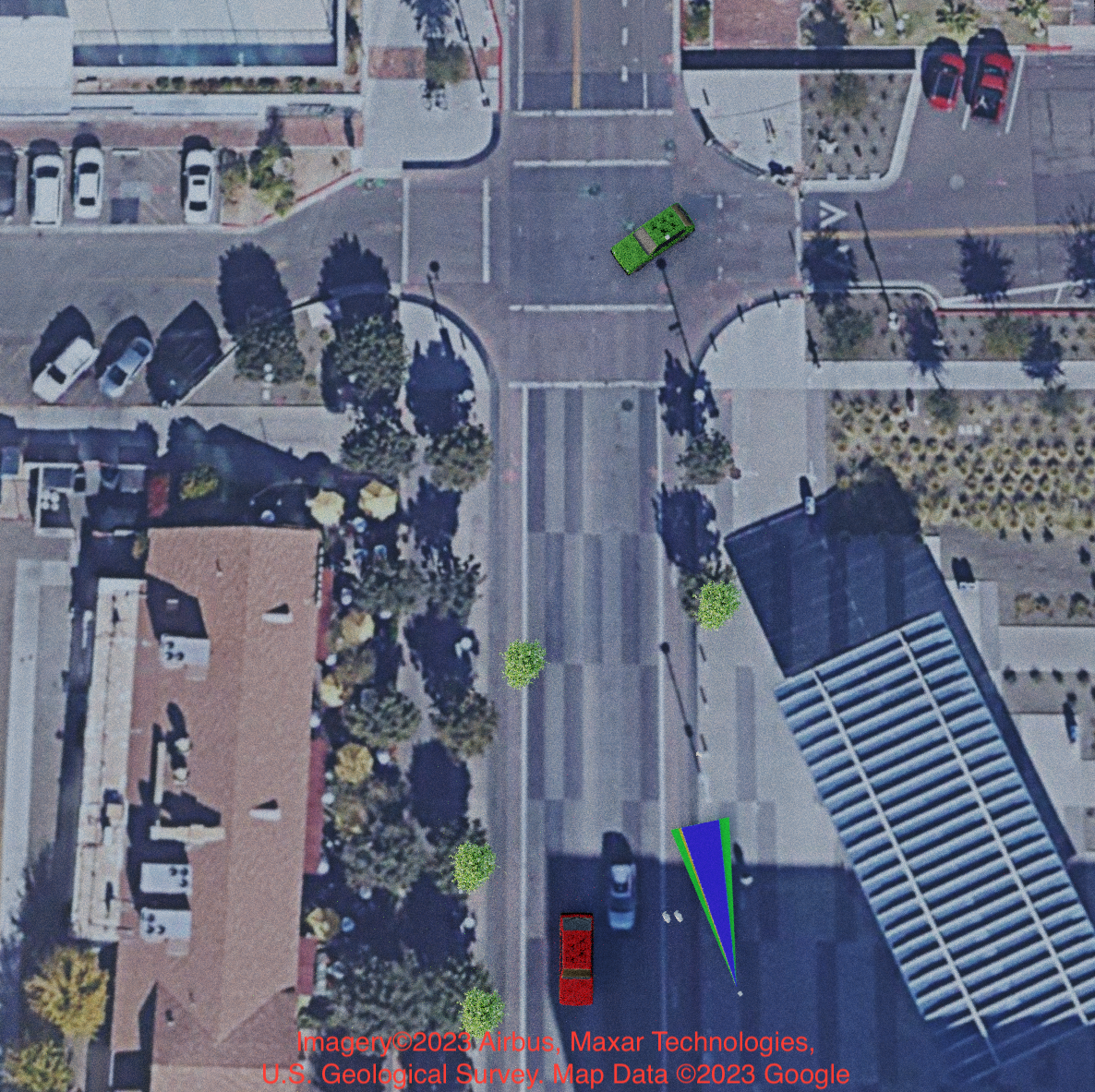}}}%
    \quad \hspace*{-0.9em}
    \subfloat[\centering Scenario 34]{{\includegraphics[width=0.24\textwidth,height=4cm]{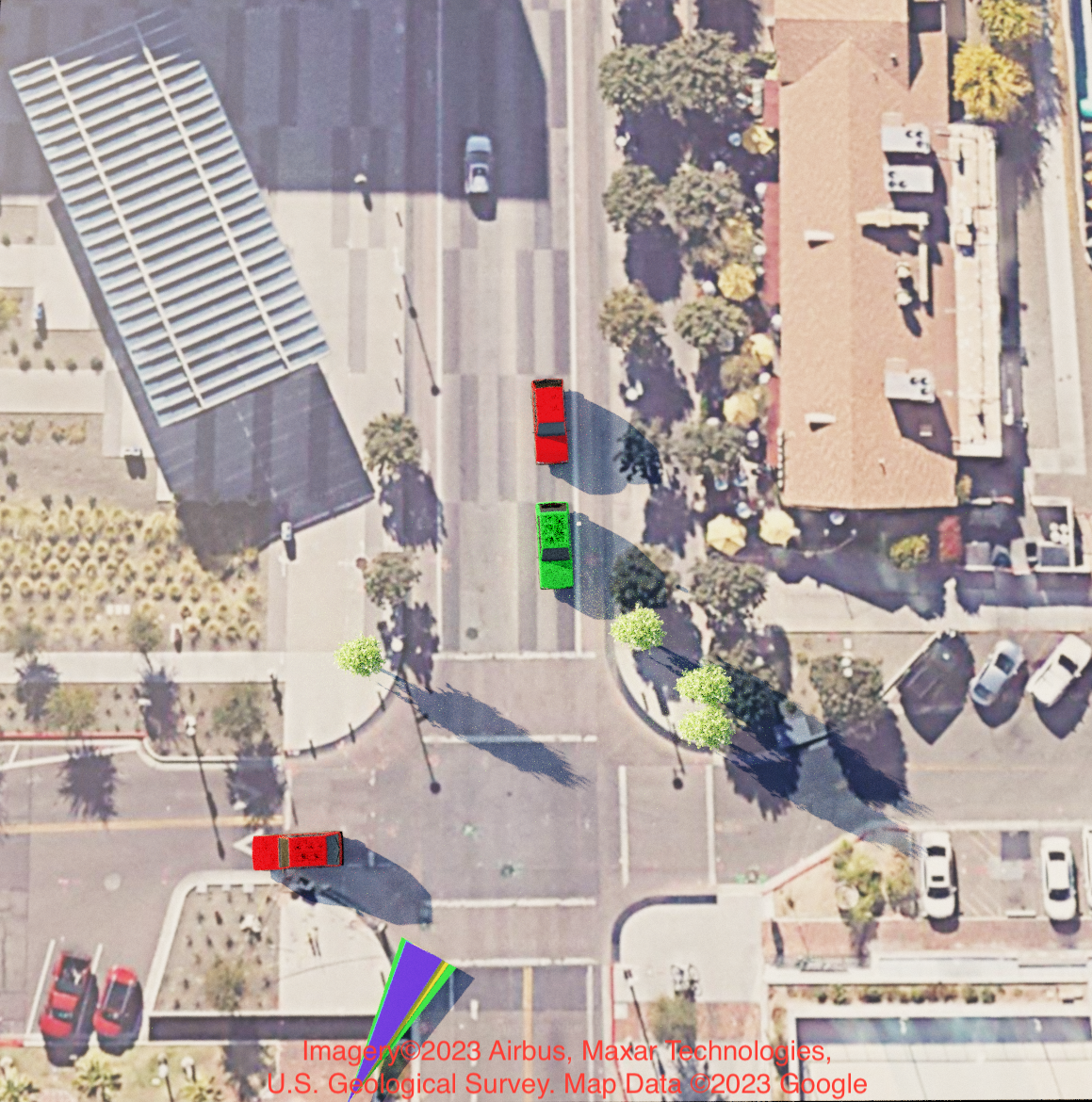}}}%
    \caption{Visualization of Digital Twins. 3D object models introduced only for presentation purposes. Floor maps created using Google Maps Imagery$\copyright$2023 Airbus, Maxar Technologies, U.S. Geological Survey. Map Data $\copyright$2023 Google}%
    \label{fig:DT_visualizaation}%
\vspace{-0.5cm}
\end{figure*}

%% file: sections/3_digital_twin.tex
\section{Digital Twin} \label{sec:digital_twin}
In this section we describe our digital twin pipeline that consists of three primary components: 3D reconstruction, channel simulation and an optional adaptation stage.

\subsection{3D reconstruction}
\begin{figure}
     \includegraphics{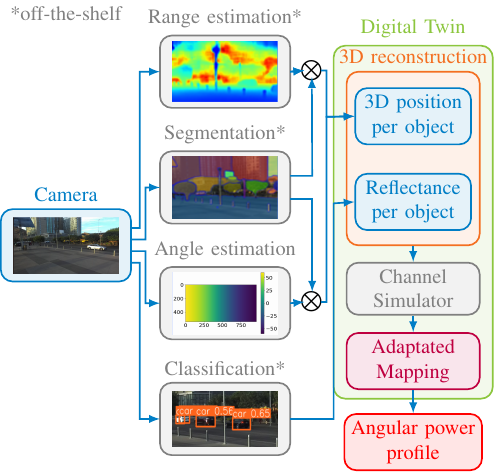}
     \caption{Proposed pipeline: Leveraging advanced computer vision techniques to create a \ac{DT} that estimates the angular power profile.}
\vspace{-2mm}  
    \label{fig:dt flow}
\end{figure}

To create our 3D model in an automated fashion without manual intervention, we leverage a combination of prominent techniques from the field of computer vision:
\begin{itemize}
\item Monocular depth estimation via PixelFormer\cite{agarwal2023attention} allows us to estimate depth information per pixel from a single image,
\item Pinhole transformation with known camera parameters to estimate a 3D direction vector per pixel location,
\item Semantic segmentation via Segment Anything\cite{kirillov2023segment} provides us with 2D pixel masks for each object in the scene,
\item Object detection via YOLO (You Only Look Once) \cite{redmon2016you} provides a class label and a 2D bounding box for each object in the scene.
\end{itemize}

First, we compute the 3D position of each pixel based on the monocular depth estimation (range) and pinhole coordinate transformation (azimuth, elevation). Then, we use semantic segmentation to identify the pixels of each object in the scene. The \textbf{3D position} of the object is the mean 3D position of its corresponding pixels. 

To estimate the \textbf{reflectance} of the object, we use the object detection system. We find the 2D bounding box with the largest \ac{IoU} with the object's 2D pixel mask. Then, we map its class label to a scalar reflectance value (tree: $0.3$, car: $1.0$, pole: $0.6$).

Our \textbf{3D representation} models each reflector as a \emph{point reflector} with a scalar reflectance. Fig.~\ref{fig:dt flow} provides an overview of the algorithm. 

All of the techniques used in this pipeline are off-the-shelf solutions, requiring no further training.
Fig.~\ref{fig:DT_visualizaation} shows the four different scenarios reconstructed from the camera view. 

We identify the \ac{UE} object by selecting the object closest to the \ac{UE} \ac{DGPS} position.

\subsection{Channel simulator}
Since our 3D representation has reduced material property and geometry information, we customize our channel simulator accordingly. 
Our angular power profile calculation, shown in Alg.~\ref{alg:3d_channel_sim}, is based on the assumptions that, in mmWave scenarios, the main propagation effects are first-order reflections and blockages.
\begin{algorithm}[t]
\caption{Channel simulation}\label{alg:3d_channel_sim}
\begin{algorithmic}
\algrenewcommand\algorithmicrequire{\textbf{Input:}}
\algrenewcommand\algorithmicensure{\textbf{Calculate:}}
\Require\\
\begin{itemize}
\item Antenna position $\mathbf{p}_\text{UE} \in \mathbb{R}^3$ derived from GPS, objects position $\mathbf{p}_i \in \mathbb{R}^3$, both in camera reference frame.
\item  Azimuth of objects ${\alpha}_i \in \mathbb{R}$ and of UE $\alpha_\text{UE} \in \mathbb{R}$
\item  Estimated reflectance of objects $\gamma_i \in [0, 1]$
\item  Wavelength $\lambda$ for the given deployment frequency
\item  Receiver hardware's angular resolution $\alpha_\text{HW}=\SI{10}{\degree}$ and angular impulse response ${s}\left(\alpha\right)=\text{sinc}(\frac{\alpha}{\alpha_\text{HW}})$ 
\end{itemize}
\Ensure
\State 1. Pathloss for object i (and UE) path:\newline \hspace*{2em} $\beta_i = \left(\frac{\lambda}{4\pi(\norm{\mathbf{p}_i-\mathbf{p}_\text{UE}}+\norm{\mathbf{p}_\text{i}})}\right)^2; \beta_{\text{UE}} = \left(\frac{\lambda}{4\pi \norm{\mathbf{p}_{\text{UE}}}}\right)^2$
\State 2. Angular power profile:\newline \hspace*{2em} $\Tilde{r}_{\alpha}\left(\alpha\right) =  \beta_{\text{UE}}\delta\left({\alpha-\alpha_\text{UE}}\right) + \sum_{i=1}^{N_\text{objects}}\gamma_{i}\beta_{i}\delta\left({\alpha-\alpha_i}\right)$
\State 3. Simulate limited receiver hardware resolution:\newline \hspace*{2em} ${r}_{\alpha,\text{DT}}\left(\alpha\right) = \left(\Tilde{r}_{\alpha} * {s}\right)\left(\alpha\right)$
\end{algorithmic}
\end{algorithm}
In this calculation, the standard propagation equations (e.g. path-loss, angle of arrival) are used to calculate a propagation path for all objects and the \ac{LoS} path, then the limited receiver hardware resolution is simulated\footnote{Angular resolution estimated through anechoic chamber measurements.} . The angular power profile is sampled in \SI{1}{\degree} steps with coverage of $N_\text{angles}=180$. The result can be represented in the angular domain by a vector $\mathbf{r}_{\alpha,\text{DT}}$.
  
\subsection{Adaptation}
Due to the simplified channel simulation, we could exclude some critical components, e.g. scattering and diffraction, causing a significant gap between simulation and real measurements. 

In order to close this gap, we use a linear angular mapping $\mathbf{M} \in \mathcal{R}^{180\times180}$.
The values of the matrix $\mathbf{M}$ are set to minimize the mean squared error between the mapped angular power profile and the ground truth computed from real measurements. This is achieved by optimizing the objective function (\ref{eq:sim2real_loss}) via \ac{SGD} with a batch-size of 256 and a learning rate of 1e-3. 
\begin{equation}\label{eq:sim2real_loss}
\mathcal{L}=   \frac{1}{N_{\text{batch}}}\sum^{N_{\text{batch}}-1}_{n=0} \norm{\mathbf{M}\mathbf{r}_{\alpha,\text{DT},n} - \mathbf{r}_{\alpha,n}}^2
\end{equation}

%% file: sections/4_downstream.tex
\section{Downstream tasks} 
\label{sec:downstream}
To demonstrate the benefits of a \ac{DT}-based solution over end-to-end approaches, we will consider the task of beam acquisition. We simplify the problem statement as follows.

In a regular setting, the \ac{BS} has no information about the whereabouts of the \ac{UE} or any other environmental information, so it has to sweep its \ac{L1} beams. The \ac{UE} then reports the received power values to the \ac{BS}. Based on these values, the \ac{BS} selects an L1 beam to further refine. This is then done by sweeping the corresponding \ac{L2} beams with a finer beam pattern. 

To minimize the beam management overhead, we evaluate two tasks. First, we have \textbf{\ac{L1} prediction} where  we can skip the sweeping of the \ac{L1} beams by predicting which one to refine further. If this is done accurately, only the corresponding \ac{L2} beams need to be swept next. 

Next, we have \textbf{\ac{L2} prediction} where we can skip the sweeping of both the \ac{L1} and \ac{L2} beams by predicting a single or a few \ac{L2} beams to measure the receive power for. In this case, the requirement is to minimize the energy loss compared to the beam with maximum received power.

In our solution, for both tasks, we determine the predicted beams by using Eq. \ref{eq:powerrecon} to estimate the received power for all beams in the given \ac{L1}/\ac{L2} codebook, and selecting the top-$K$ with the highest value. The number of selected beams ($K$) is set differently in each of the benchmarks to adjust for task difficulty.

We compare this solution with a \ac{GPS}-only \ac{LoS} baseline, and a learned end-to-end solution, both described in detail in the experiments section.

%% file: sections/5_experiments.tex
\section{Experiments} \label{sec:experiments}
\input{tables/results_sota.tex}
\subsection{Metrics}
For the experiments comparing to state-of-the-art, we use the \ac{DBA} score suggested by the DeepSense6G beam prediction challenge, and described in detail in \cite{avatar2022}. We note that the \ac{DBA} score accounts only for the \emph{proximity} of the predicted indices to the best ones, not accounting for received power, for example from reflectors. Since we wanted to analyze the impact on the RF communication system, we propose different metrics for our downstream tasks.

For the task of \textbf{L1 prediction} we predict the L1 beam that will be refined by L2 beam sweep in the next stage, so we need to maximise L1 prediction accuracy, where accuracy is defined as: 
 \begin{equation}
L_{1,\text{acc}} = \frac{1}{N}\sum_{n=0}^{N} \begin{cases}
    1, & \text{if } c_n \in \hat{\mathbf{c}}_n\\
    0,              & \text{otherwise},
\end{cases} 
\end{equation}
where $c_n$ is the best beam index of sample $n$ and $\hat{\mathbf{c}}_n$ the predicted beam indices. 

For the task of \textbf{L2 prediction}, the primary objective is to find the beam that maximises the received power. In other words, the used metric should proportionally reflect the energy loss, not just the accuracy of the system. For this reason, the main metric is the received power loss:
\begin{equation}
    P_\text{loss} = 20\log_{10}{\frac{1}{N}\sum_{n=0}^{N} \left(\frac{  y_{c,n}}{\max \mathbf{y}_{\hat{c},n}}\right)},
\end{equation}
where $\mathbf{y}_{\hat{c},n}$ is the received power of the top-$K$ predicted beams and $y_{c,n}$ is the received power of the best beam. 

As analysing the system behavior in \ac{LoS} and \ac{NLoS} cases separately can lead to additional insights, we use 50th and 95th percentile of the power loss values to represent \ac{LoS} and \ac{NLoS} cases, respectively.

While we analyze all scenarios, it is important to note the difference between them. Scenario 32, 33 and 34 are considered \textbf{seen} scenarios, ie., they are also part of the training set. On the other hand, Scenario 31 is \textbf{unseen}, i.e., not part of the training set, closer reflecting a real-world deployment setting. The overall DBA score used in the challenge reflects this difference, by assigning $1/2$ weight for the unseen, and $1/6$ to each of the seen scenarios.

\subsection{Baselines}
In order to analyse the behavior of the proposed system, we utilize both a \ac{GPS}-only system that is only expected to work in \ac{LoS} conditions and a more flexible neural network based end-to-end approach.

Our \ac{GPS} solution (referred to as \textbf{GPS-LoS}) simply chooses the beam with the maximum beam profile value at the \ac{UE} azimuth, assuming \ac{LoS} conditions.

As the state-of-the-art approach\cite{avatar2022} has the downstream task built into its design, we needed to introduce an alternative baseline that estimates the angular power profile directly, allowing for a flexible application to new downstream tasks. Such a system can serve as substitute for \ac{DT}-creation and channel simulation stages jointly, allowing us to have direct comparison with a \ac{DT}-based approach.
\begin{figure}
     \includegraphics{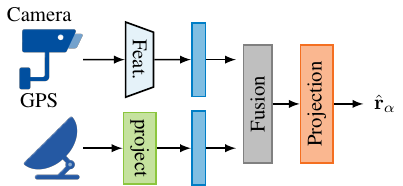}
     \caption{The proposed end-to-end \ac{NN} architecture, relying on standard off-the-shelf feature extractor (ResNet-18) and a projection head.}
\vspace{-2mm}  
    \label{fig:model_arch_1}
\end{figure}
For this purpose, we utilize a multi-modal \ac{NN}, shown in Fig.~\ref{fig:model_arch_1}.
 To extract features from images, we employ the ResNet-18 architecture
for which we substitute the final projection layer with a linear projection to a 180-dimensional space. Regarding the \ac{GPS} modality, we represent the \ac{UE} position as an azimuth angle, then use a projection layer to a 180-dimensional space. The extraction step is then followed by feature fusion concatenation and a two-layer \ac{MLP}. The final output is represented by a 180-dimensional vector matching the angular power profile dimension. 
As supervision, we utilize a \ac{MSE} angle reconstruction loss:
\begin{equation}\label{eqn:loss}
\mathcal{L}(\mathbf{
    \hat{r}}_{\alpha},\mathbf{r}_{\alpha}) = \frac{1}{N_{\text{batch}}}\sum^{N_{\text{batch}}-1}_{n=0}\norm{\mathbf{
    \hat{r}}_{\alpha,n}-\mathbf{r}_{\alpha,n}}^2.
\end{equation} 
The training is done with a learning rate of $0.001$ using the ADAM optimizer. 
We will refer to our neural network based solution as the \textbf{end-to-end} approach. Table \ref{tab:sota-results} shows that our proposed end-to-end baseline is on par with the state-of-the-art (and even outperforms it).

\subsection{DeepSense 6G beam prediction}
Table \ref{tab:sota-results} demonstrates the state-of-the-art results on the challenge test dataset. Our DT approach with and without adaptation are referred to as \textbf{DT-Adapt} and \textbf{DT}, respectively. 

We demonstrate that we improved the overall DBA score by infusing propagation expertise as well as learnable features. We would also like to highlight that our \ac{DT} solution \emph{without adaptation} is also highly competitive and vastly outperforms the state-of-the-art, even without any training/measurements involved. Our DT-Adapt solution improves performance further, even though it only relies on a small calibration set, not the full training set, suggesting a reduced requirement on large-scale data collection.

The end-to-end solution is showing strong performance on "seen" scenarios, as its number of trainable parameters allow it to overfit. Supporting our hypothesis is the end-to-end solutions performance in "unseen" cases, where even our parameter-free solution exhibits stronger generalization ability.

\subsection{Downstream tasks}
To demonstrate the benefit of using a \ac{DT}-based solution as opposed to the an end-to-end approach, we evaluated their performance on both of the downstream tasks described in Section \ref{sec:downstream}. 
The results on \textbf{L2 prediction} can be seen in Table \ref{tab:results_l2}, with the observed standard deviations (introduced by the stochastic learning process) computed from 10 runs. In the \ac{LoS} cases, while the task is easy and all of the tested methods perform well, the adapted \ac{DT} solution provides perfect predictions. 

In the \ac{NLoS} setting, in the "unseen" Scenario 31, the non-adapted \ac{DT} solution shows stronger generalization compared to the end-to-end approach. However, after adaptation, the energy loss is radically decreased, outperforming all of the approaches.

In the \ac{NLoS} setting, in the "seen" Scenarios 32 and 33, the adapted \ac{DT} also outperforms the end-to-end method, though less radically. For the case on Scenario 34, the \ac{DT}-based methods underperform compared to the end-to-end case, which requires further analysis. One possible explanation could be, that Scenario34 was measured throughout the night time. However, the overall system does provide an advantage over the GPS-LoS solution.

\input{tables/results_l1}
For the task of \textbf{L1 prediction}, the results can be seen in Table \ref{tab:results_l1}, that demonstrates that the \ac{DT} solution is on par with the end-to-end approach \emph{without requiring any RF measurements as training data}.
\input{tables/results_l2.tex}
\subsection{Improved explainability}
\begin{figure}
     \includegraphics{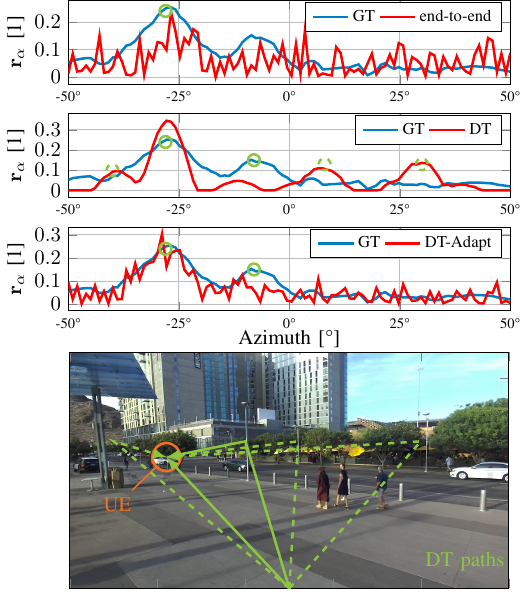}
        \vspace{-4mm}
     \caption{Example for a reconstructed angular power profile comparing to the camera image and identifying useful reflectors. Dashed paths are correctly removed during the adaptation phase, while the dominant paths can be seen in the reconstructed profile. (Scenario 32, Sample 50)}
\vspace{-2mm}  
    \label{fig:explainability}
\end{figure}
Fig. \ref{fig:explainability} demonstrates the angular power profiles reconstructed by the different approaches.

The profile reconstructed by the end-to-end solution demonstrates no clear visible correspondence to the ground truth (GT). This is a strong example for the lack of explainability of standard \ac{NN}s.

Our \ac{DT}-reconstructed profile, however, shows clear correspondence with the objects visible on the camera image (marked with green circles). It is worth noting that the \ac{DT} identified reflectors that should not be able provide a reflecting path to the \ac{UE}, as the \ac{UE} vehicle's frame is expected to block those paths (see paths reflected from poles on the right), or the normal vector of the surface was not appropriate (see path reflected from column on left). These phenomena are not modeled in our proposed 3D representation.

A solution to this problem is provided by our proposed adaptation technique, as it keeps the \ac{LoS} path, and suppresses the paths introduced by our simplified 3D representation.

We note that the dataset sample used for this analysis was not cherry-picked, we used the first example with multiple reflections in the scene.

%% file: tables/results_sota.tex
\begin{table*}[]
\vspace{2mm}
\centering
\caption{Results: Comparison to DeepSense6G challenge}
\vspace{-2mm}
\label{tab:sota-results}
\begin{tabular}{|c|c||c|c|c|c|c|c|c}
\hline
\multicolumn{1}{|c|}{\multirow{2}{*}{Method}}   & \multicolumn{1}{c||}{\multirow{2}{*}{Model trained on}}     
& \multicolumn{1}{c|}{\multirow{2}{*}{\begin{tabular}[c]{@{}c@{}}DBA Scenario31$\uparrow$\\ \textbf{Unseen}\end{tabular}}}&  \multicolumn{1}{|c|}{\multirow{2}{*}{\begin{tabular}[c]{@{}c@{}}DBA Scenario32$\uparrow$\\ Seen\end{tabular}}}   &  \multicolumn{1}{c|}{\multirow{2}{*}{\begin{tabular}[c]{@{}c@{}}DBA Scenario33$\uparrow$\\Seen\end{tabular}}}    &  \multicolumn{1}{c|}{\multirow{2}{*}{\begin{tabular}[c]{@{}c@{}}DBA Scenario34$\uparrow$\\Seen\end{tabular}}} & \multicolumn{1}{c|}{\multirow{2}{*}{\begin{tabular}[c]{@{}c@{}}DBA Overall$\uparrow$\\ \end{tabular}}} \\  
& & & & & & \\ \hline\hline
Avatar\cite{avatar2022}  &  train  + calibration sets & 0.65 &  0.70 &  0.85 &  0.71 &  0.72\\ \hline
end-to-end            &  train  + calibration sets & 0.80$\pm$0.015   & 0.95$\pm$0.021 & 0.94$\pm$0.014 & 0.90 & 0.87\\ \hline
DT                       &  not trained & 0.90 &  0.95 &  0.93 &  0.81 &  0.90  \\ \hline
DT-Adapt                   &  calibration set only & \textbf{0.93 $\pm$ 0.01}  &  0.95 $\pm$ 0.00 &  0.930 $\pm$ 0.00  &  0.833 $\pm$ 0.01 & \textbf{0.92}  \\ \hline
\end{tabular}
 \vspace{-5mm}
\end{table*}

%% file: tables/results_l1.tex
\newcolumntype{g}{>{\columncolor{lightgreen!20}}c}
\begin{table}[H]
\centering
\caption{Results on L1 prediction benchmark}
\vspace{-1mm}
\label{tab:results_l1}
\centering
\begin{tabular}{|c|c|c|c|c|c|c|c|c|c|c|c}
 \hline
\multicolumn{1}{|c|}{\multirow{2}{*}{Method}}       & \multicolumn{1}{c|}{\multirow{2}{*}{Requires Data}}  & \multicolumn{2}{c|}{L1 Accuracy$\uparrow$}            \\ \cline{3-4}
             &  & \multirow{1}{*}{Top-$1$} & \multirow{1}{*}{Top-$2$} \\ \hline \hline 
GPS-LoS      & No &\SI{77.1}{\percent}&\SI{98.8}{\percent}\\ \hline
end-to-end     & Yes &\SI{86.3}{\percent}&\SI{99.9}{\percent}\\ \hline
DT           & No &\SI{85.8}{\percent}&\SI{99.1}{\percent}\\ \hline
\end{tabular}
 \vspace{0mm}
\end{table}

%% file: tables/results_l2.tex
\begin{table*}[]
\vspace{2mm}
\caption{Results on L2 prediction benchmark}
\vspace{-2mm}
\label{tab:results_l2}
\centering
\begin{tabular}{|c|c|c|c|c|c|c|c|c|c|c|c|c|c|c}
 \hline
\multicolumn{1}{|c|}{\multirow{3}{*}{Method}}         & \multicolumn{1}{c|}{\multirow{3}{*}{Scene}}    & Seen     & \multicolumn{6}{c|}{Power Loss L2 [dB]$\downarrow$}                  \\ \cline{4-9} 
& & in & \multicolumn{3}{c}{50\% (LoS cases)}  & \multicolumn{3}{|c|}{95\%  (NLoS cases)} \\ \cline{4-9}
& &     training    & Top-$1$   & Top-$2$   & Top-$3$   & Top-$1$   & Top-$2$   & Top-$3$   \\ \hline \hline
GPS-LoS  &\multirow{4}{*}{31} &\multirow{4}{*}{\textbf{Unseen}}&0.40& 0.35 & 0.26& 3.03& 2.78  &2.74\\ \cline{1-1}\cline{4-9} 
end-to-end & &&0.34$\pm$0.1&0.0$\pm$0.0&0.0$\pm$0.0&9.68$\pm$1.3&7.15$\pm$0.9&6.41$\pm$0.6\\ \cline{1-1}\cline{4-9} 
DT       & && 0.42& 0.0 & 0.0& 6.10& 4.76  &4.11\\ \cline{1-1}\cline{4-9} 
DT-Adapt & &&0.00$\pm$0.0&0.0$\pm$0.0&0.0$\pm$0.0&$
\mathbf{0.96\pm0.1}$&$\mathbf{0.63\pm0.1}$&$\mathbf{0.41\pm0.1}$\\ \hline \hline
GPS-LoS  &\multirow{4}{*}{32} &\multirow{4}{*}{Seen}& 0.35& 0.34 & 0.33& 3.16 & 2.92  &2.90\\ \cline{1-1}\cline{4-9} 
end-to-end & &&0.19$\pm$0.0&0.0$\pm$0.0&0.0$\pm$0.0&2.27$\pm$0.3&0.86$\pm$0.3&0.63$\pm$0.2\\ \cline{1-1}\cline{4-9} 
DT       & && 0.0& 0.0 & 0.0& 3.11& 2.41  &1.62\\ \cline{1-1}\cline{4-9} 
DT-Adapt & &&0.00$\pm$0.0&0.0$\pm$0.0&0.0$\pm$0.0&1.15$\pm$0.1&0.75$\pm$0.1&0.55$\pm$0.1\\ \hline \hline
GPS-LoS  &\multirow{4}{*}{33} &\multirow{4}{*}{Seen}& 0.52& 0.51 & 0.50& 3.70 & 3.62  &3.60\\\cline{1-1}\cline{4-9} 
end-to-end & &&0.0$\pm$0.0&0.0$\pm$0.0&0.0$\pm$0.0&2.21$\pm$0.7&1.15$\pm$0.3&0.82$\pm$0.1\\ \cline{1-1}\cline{4-9} 
DT       & && 0.40& 0.07 & 0.0& 3.95& 1.18  &0.93\\ \cline{1-1}\cline{4-9} 
DT-Adapt & &&0.00$\pm$0.0&0.0$\pm$0.0&0.0$\pm$0.0&1.25$\pm$0.2&0.63$\pm$0.1&0.43$\pm$0.1\\ \hline \hline
GPS-LoS  &\multirow{4}{*}{34} &\multirow{4}{*}{Seen}&1.82 & 1.24 & 1.21& 7.18 & 5.96  &5.76 \\ \cline{1-1}\cline{4-9} 
end-to-end & &&0.0$\pm$0.0&0.0$\pm$0.0&0.0$\pm$0.0&1.21$\pm$0.4&1.01$\pm$0.1&0.50$\pm$0.1\\ \cline{1-1}\cline{4-9} 
DT       & &&0.77& 0.46 & 0.12& 7.87& 6.79 &5.94\\ \cline{1-1}\cline{4-9} 
DT-Adapt & &&0.20$\pm$0.1&0.0$\pm$0.0&0.0$\pm$0.0&3.31$\pm$0.1&2.78$\pm$0.1&2.2$\pm$0.1\\ \hline
\end{tabular}
 \vspace{-4mm}
\end{table*}

%% file: sections/6_conclusions.tex
\section{Conclusion} \label{sec:conclusions}
In conclusion, we propose a pipeline of \ac{DT} creation and channel simulation including a learnable adapatation stage, which is versatile for tackling mmWave beam management downstream tasks. The described pipeline is practical and relies only on a single camera co-located with the \ac{BS}, and UE position information. We demonstrate the performance benefits compared to end-to-end approaches on the downstream task of beam acquisition, using the real-world dataset of the DeepSense6G challenge.

Due to the chosen 3D representation, an added benefit of our approach over end-to-end solutions is that it is extendable, for example to blockage prediction through the trajectory prediction of objects in the scene. Since our channel simulator is simple and differentiable, it allows for further extensions, for example to codebook learning. We leave the exploration of such approaches to future work.